\documentstyle[twoside,epsf]{article}

\input ibvs2.sty

\begin{document}

\IBVShead{5865}{1 December 2008}

\IBVStitletl{Long-term optical observations of the}{ Be/X-Ray binary system V0332+53}

\IBVSauth{Kizilo\u{g}lu, \"U.; Kizilo\u{g}lu, N.; Baykal, A.; Yerli S. K.; \"Ozbey, M.}

\IBVSinst{Physics Dept., Middle East Technical University, Ankara 06531, Turkey \\ e-mail: umk, nil, altan, sinan, mehtap@astroa.physics.metu.edu.tr}

\SIMBADobjAlias{BQ Cam}{V0332+53}

\begintext

The Be/X-ray binary V0332+53 has an orbital period of 34.25\,d with an
eccentricity of 0.31 (Stella et al.~1985). The optical counterpart of this
system, BQ~Cam, is an O8-9Ve star at a distance of about 7\,kpc, showing
H${\alpha}$ line emission (Negueruela et al.~1999). This emission is related
to the circumstellar disk around the optical star.

Three optical brightening of BQ~Cam have been detected. Two of them were
reported by Goranskij (2001), one was in 1983 and the other was in 1989. The
third one was reported by Goranskij and Barsukova (2004) in the beginning of
2004. About 300 days later, Swank et al.~(2004) informed the first All Sky
Monitor detection (on the Rossi X-Ray Timing Explorer (RXTE)) of the November
2004 X-ray outburst. The previous two optical brightenings were also
accompanied by X-ray outbursts.

Recently, Krimm et al.~(2008) reported a new X-ray activity starting at
MJD~54\,756 detected by
Swift/BAT\footnote{http://swift.gsfc.nasa.gov/docs/swift/results/} hard X-ray
transient monitor. Hsiao et al.~(2008) obtained an optical spectrum at
MJD~54\,761 in which the H${\alpha}$ emission line showed P-Cygni profile with
FWHM $\sim$12\,\AA.

We have been monitoring the binary system V0332+53 since 2004 using the 45\,cm
ROTSEIIId telescope (Robotic Optical Transient
Experiment)\footnote{http://www.rotse.net} and RTT150 (Russian-Turkish 1.5 m
Telescope)\footnote{http://www.tug.tubitak.gov.tr} located at
Bak{\i}rl{\i}tepe, Antalya, Turkey. ROTSEIII telescopes which operate without
filters were described in detail by Akerlof et al.~(2003). Details on the
reduction of the data were described in Baykal et al.~(2005) and
K{\i}z{\i}lo\u{g}lu et al.~(2005). The reference stars for differential
photometry were listed in a previous study of Baykal et al.~(2005).

In our previous study (Baykal et al.~2005), we presented part of the optical
light curve during the giant 2004 X-ray outburst. In this study we report on
the long-term variability of the Be/X-ray binary system V0332+53 up to the
present date. The differential optical light curve and X-ray light curve of
Be/X-ray binary system V0332+53 are shown in Fig.~1. X-ray light curve was
obtained from RXTE/ASM web site\footnote{http://xte.mit.edu}.

A fading of 0.2\,mag occurs in the light of BQ Cam after MJD~53\,400. On the
onset of the fading trend, the Type II X-ray outburst comes to an end. The
X-ray activity ends accompanied by the fading of magnitudes. The fading in the
light curve of BQ~Cam could be due to a decrease in the density or in the size
of the circumstellar disk. After MJD~53\,600 the system brightened again but
did not reach its previous value observed before the giant 2004 X-ray activity
until about MJD~54\,700.

\IBVSfig{4.0in}{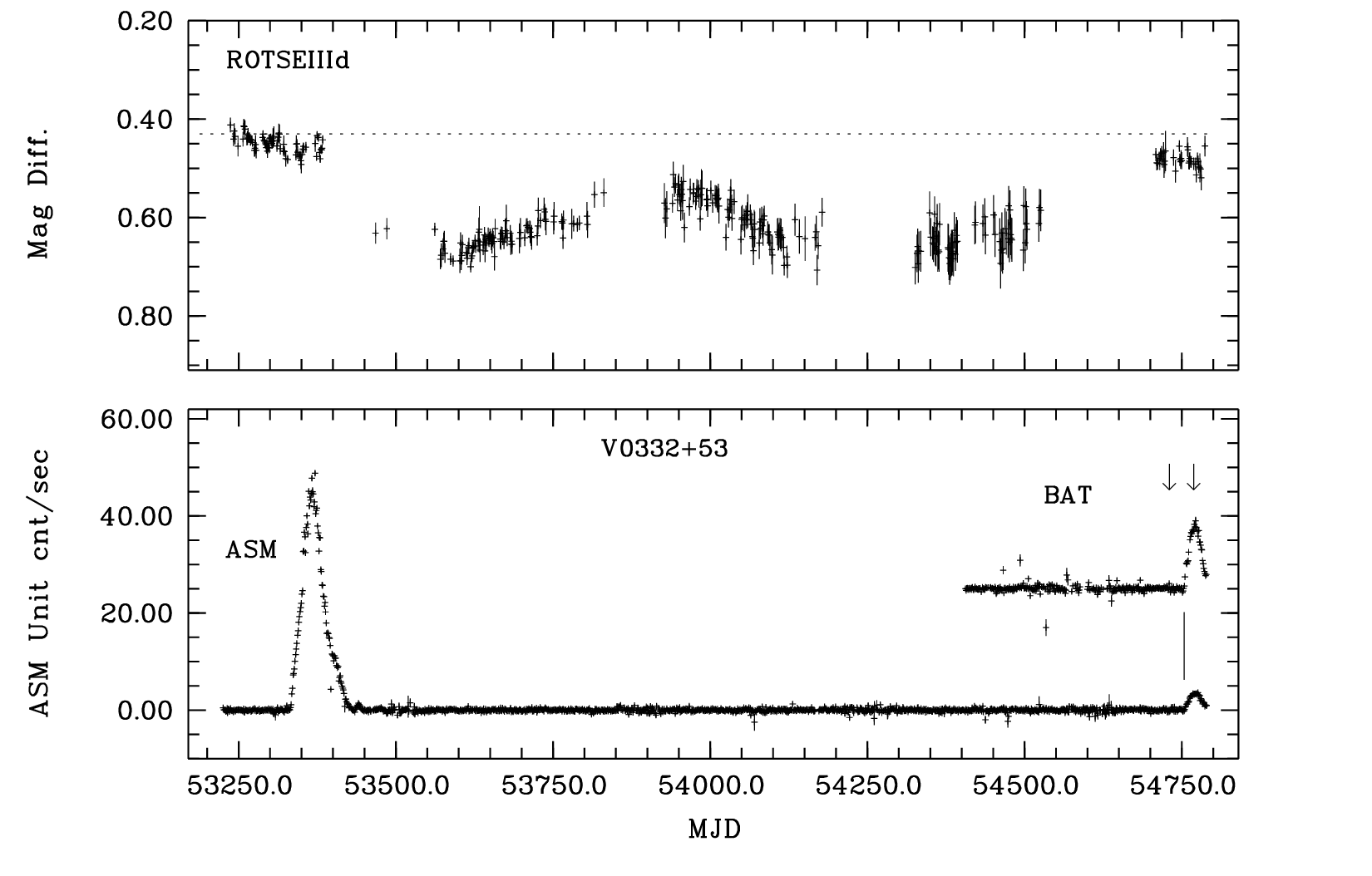}{ROTSEIIId daily averaged differential light curve
(upper panel) and X-ray light curve (lower panel) of the Be/X-ray system
V0332+53 (MJD = JD $-$ 2400000.5). Daily averages of RXTE/ASM 5.0-15.0 keV
band light curve and 15-50\,keV SWIFT/BAT light curve (properly scaled and
shifted) are shown. Vertical line represents PAP and arrows denote
spectroscopic observation times.}

\IBVSfig{3.5in}{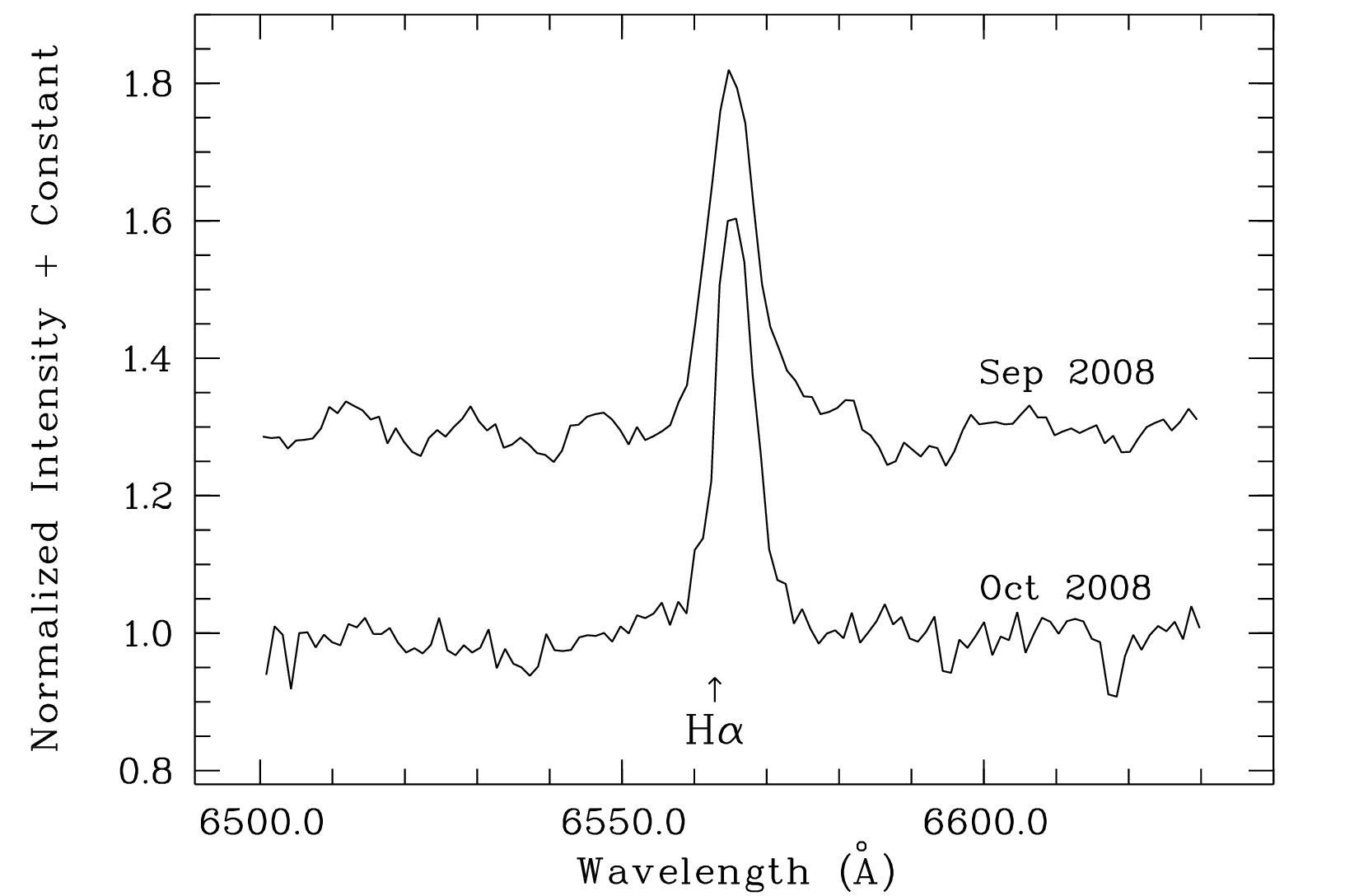}{H${\alpha}$ profiles observed on Sep 21 and Oct
29, 2008, before and during the X-ray activity.}

We presented optical spectroscopic observations obtained before (at
MJD~54\,730) and during (at MJD~54\,768) the new X-ray activity reported by
Krimm et al.~(2008). The spectroscopic observations were performed with the
RTT150 telescope using the medium resolution spectrometer TFOSC (T\"UB\.ITAK
Faint Object Spectrometer and Camera). The camera is equipped with a
$2048\times2048$, 15$\mu$ pixel Fairchild 447BI CCD. We used grism G8
(spectral range 5800-8300\,\AA) with average dispersion of
$\sim$1.1\,\AA~pixel$^{-1}$. The reduction and analysis of spectra were made
by using MIDAS\footnote{http://www.eso.org/projects/esomidas/} and its
packages: Longslit context and ALICE.

The observed H${\alpha}$ line profiles (Fig.~2) were single-peaked and almost
symmetric. Measurements of H${\alpha}$ emission lines were made by fitting a
Gaussian profile. For each spectrum the measured value of the equivalent width
(EW) and full width at half maximum (FWHM) are given in Table 1. The EW and
FWHM values for the present two H${\alpha}$ emission profiles are almost the
same. The calculated EW value of $\sim4.4$\,\AA for both profiles is less than
the measured value of 10\,\AA which was obtained by Masetti et al.~(2005) at
MJD~53\,377. It should be noted that the Be disk was denser at that time.
According to the present data, the disk is less dense and the system has
almost reached the previous brightness observed before the giant X-ray flare.

The present EW values are found to be similar to the ones observed during the
fading of infrared magnitudes of Negueruela et al.~(1999). We did not confirm
the  result of Hsiao et al.~(2008) since our detection showed single peaked
H${\alpha}$ emission line (obtained 7 days later than their observations). In
addition to this, the present FWHM was weaker by a factor of 2.

The H${\alpha}$ emission lines were found to be red-shifted by $\sim$140\,km/s
which were larger than that of Corbet et al.~(1986), who found a blue-shift of
$\sim$65 km/s in H${\alpha}$ line and related this to V/R variability seen in
Be type stars. In the present study, quite symmetric H${\alpha}$ line profiles
do not represent a perturbation in the disk. Because of the low inclination of
this system, it is also possible that no variability is seen.

Okazaki and Negueruela (2001) pointed out the possibility of disc truncation
by the neutron star which was not close to the mean critical Roche Lobe radius
at periastron for the binary system V0332+53 since this system showed no Type
I X-ray outburst for a long period of time. According to them, to have a
temporary Type I X-ray outbursts, Be disk should be strongly disturbed. But,
the H${\alpha}$ emission line profile obtained during the 2008 Type I X-ray
outburst does not show any variability which would indicate a disturbed disk.
The line is quite symmetric.

\begin{table}[!bbb]
\normalsize
\begin{center}
{\normalsize {\bf Table 1.} H$\alpha$ line profiles.}
\vskip 3mm
\begin{tabular}{cccc}
\hline
Date & MJD & EW (\AA)  & FWHM (\AA)   \\
\hline
Sep 21, 2008     & 54730.0796 & 4.44 $\pm$ 0.13 & 7.89 $\pm$ 2.05 \\
Oct 29, 2008     & 54768.8644 & 4.37 $\pm$ 0.15 & 6.57 $\pm$ 1.52 \\
\hline
\end{tabular}
\end{center}
\end{table}

We suggest that  brightening of the disk after MJD~54\,700 may be due to the
precession of the disk. When the disk is toward the periastron the material in
the outer part of the disk falls on to the neutron star giving rise to the
observed 2008 X-ray outburst. The new 2008 X-ray outburst coincides with the
periastron passage (PAP) time of the neutron star (Type I outburst).  We used
the orbital period of 34.67 days and PAP time of 53367 given by Zhang et
al.~(2005).

We continue monitoring the system.

\bigskip

\textit{Acknowledgments:} This project utilizes data obtained by the Robotic
Optical Transient Search Experiment.  ROTSE is a collaboration of Lawrence
Livermore National Lab, Los Alamos National Lab and the University of Michigan
(http://www.rotse.net). We thank the Turkish National Observatory of
T\"UB\.ITAK for running the optical facilities. This study was supported by
TUG (Turkish National Observatory), T\"UB\.ITAK ( Turkish Scientific and
Technological Research Council), through project 106T040.

\references

Akerlof, C. W., Kehoe, R. L., McKay, T. A., Rykoff, E. S., Smith, D. A., et al. 2003, {\it PASP}, {\bf 115}, 132

Baykal, A., K{\i}z{\i}lo\u{g}lu, U., K{\i}z{\i}lo\u{g}lu, N. 2005, {\it IBVS}, 5615

Corbet, R. H. D., Charles, P. A., van der Klis, M. 1986, {\it A\&A}, 162, 117

Goranskij, V. P. 2001, {\it AstL}, 27, 516

Goranskij, V., Barsukova, E. 2004, {\it ATel}, No. 245

Hsiao, E. Y., et al. 2008, {\it ATel}, 1803

K{\i}z{\i}lo\u{g}lu, U., K{\i}z{\i}lo\u{g}lu, N., Baykal, A. 2005, {\it AJ}, 130, 2766

Krimm, H. A., et al. 2008, {\it ATel}, No. 1792

Masetti, N., Orlandini, M., Marinoni, S., Santangelo, A. 2005, {\it ATel}, No. 388
 
Negueruela, I., Roche, P., Fabregat, J., Coe, M. J. 1999, {\it MNRAS}, 307, 695

Okazaki, A. T., Negueruela, I. 2001, {\it A\&A}, 377, 161

Stella, L., White, N. E., Davelaar, J., et al. 1985, {\it ApJ}, 288, L45

Swank, J., remilliard, R., Smith, E. 2004, {\it ATel}, No. 349

Zhang, S., Qu, J. L., Song, L. M., Torres, D. F. 2005, {\it ApJ}, 630, L65

\endreferences

\IBVSedata{5865-t1.txt}
\IBVSdataKey{5865-t1.txt}{BQ Cam}{photometry}
\IBVSedata{5865-t2.txt}
\IBVSdataKey{5865-t2.txt}{BQ Cam}{spectrum}
\IBVSedata{5865-t3.txt}
\IBVSdataKey{5865-t3.txt}{BQ Cam}{spectrum}

\end{document}